\documentclass[journal]{IEEEtran}
\IEEEoverridecommandlockouts

\usepackage{cite}
\usepackage{amsmath,amssymb,amsfonts}
\usepackage{graphicx}
\usepackage{multirow}
\usepackage{subcaption}
\usepackage{caption}
\usepackage{float}
\usepackage{textcomp}
\usepackage{xcolor}
\usepackage{tikz}
\usepackage{pgfplots}
\usepackage{stfloats}
\usepackage{mathrsfs}
\usepackage{balance}
\usepackage{amsthm} 
\usepackage[ruled,vlined]{algorithm2e}

\usepackage{comment}
 \usepackage{booktabs}
\usepackage{algpseudocode}
\usepackage{acronym}
\usepackage[hyperfirst=true]{glossaries}
\usepackage[hidelinks]{hyperref} 
\usepackage{makecell}
\makeglossaries
\newacronym{OFDM}{OFDM}{orthogonal frequency division multiplexing}
\newacronym{OTFS}{OTFS}{orthogonal time frequency space}
\newacronym{AFDM}{AFDM}{affine frequency division multiplexing}
\newacronym{SIMO}{SIMO}{single-input multiple-output}
\newacronym{DL}{DL}{deep learning}
\newacronym{EVM}{EVM}{error vector magnitude}
\newacronym{BER}{BER}{bit error rate}
\newacronym{HSR}{HSR}{high-speed railways}
\newacronym{V2V}{V2V}{vehicle-to-vehicle}
\newacronym{V2I}{V2I}{vehicle-to-infrastructure}
\newacronym{V2X}{V2X}{vehicle-to-everything}
\newacronym{UAV}{UAV}{unmanned aerial vehicle}
\newacronym{ICI}{ICI}{intercarrier interference}
\newacronym{ISI}{ISI}{intersymbol interference}
\newacronym{IPI}{IPI}{interpath interference}
\newacronym{CFR}{CFR}{channel frequency response}
\newacronym{SISO}{SISO}{single-input single-output}
\newacronym{DoA}{DoA}{direction-of-arrival}
\newacronym{TX}{TX}{transmitter}
\newacronym{RX}{RX}{receiver}
\newacronym{CP}{CP}{cyclic prefix}
\newacronym{RMSE}{RMSE}{root mean squared error}
\newacronym{MCRLB}{MCRLB}{modified Cramér-Rao Lower Bound}
\newacronym{AWGN}{AWGN}{additive white gaussian noise}
\newacronym{SNR}{SNR}{signal-to-noise ratio}
\newacronym{MSE}{MSE}{mean squared error}
\newacronym{MLP}{MLP}{multilayer perceptron}
\newacronym{CFI}{CFI}{conditional Fisher information}
\newacronym{DD}{DD}{decision-directed}
\newacronym{ULA}{ULA}{uniform linear array}
\newacronym{DFT}{DFT}{discrete Fourier transform}
\newacronym{QAM}{QAM}{quadrature amplitude modulation}
\newacronym{MF}{MF}{matched filter}
\newacronym{MRC}{MRC}{maximum ratio combining}
\newacronym{CFAR}{CFAR}{constant false alarm rate}
\newacronym{LS}{LS}{least squares}
\newacronym{FNN}{FNN}{feedforward neural network}
\newacronym{CSI}{CSI}{channel state information}
\newacronym{zD}{zD}{zero Doppler}
\newacronym{PSD}{PSD}{power spectral density}
\newacronym{ReLU}{ReLU}{rectified linear unit}
\newacronym{LoS}{LoS}{line-of-sight}
\usepackage{etoolbox}
\usepackage{orcidlink}
\usepackage{tabularx,booktabs} %<-- Tables
\allowdisplaybreaks

\begin{document}

%\title{DL-Enhanced Multi-Antenna OFDM Receiver for 6G High-Mobility Communications}

%\title{Unlocking the Full Potential of OFDM for 6G High-Mobility Communications via Deep Learning}

%\title{Angle-Domain Processing with Deep Learning for 6G OFDM Communications with High Mobility Transceivers and Objects}

\title{6G OFDM Communications with High Mobility Transceivers and Scatterers via Angle-Domain Processing and Deep Learning}

\author{\IEEEauthorblockN{Mauro Marchese\IEEEauthorrefmark{1}, Musa Furkan Keskin\IEEEauthorrefmark{2}, Henk Wymeersch\IEEEauthorrefmark{2}, Pietro Savazzi\IEEEauthorrefmark{1}\IEEEauthorrefmark{3}} \\
\IEEEauthorblockA{\IEEEauthorrefmark{1}University of Pavia, Italy,  \IEEEauthorrefmark{2}Chalmers University of Technology, Sweden, \IEEEauthorrefmark{3}CNIT Consorzio Nazionale Interuniversitario per le Telecomunicazioni, Pavia, Italy \\
E-mail: mauro.marchese01@universitadipavia.it}
\thanks{This work is supported, in part, by the European Union under the Italian National Recovery and Resilience Plan (NRRP) of NextGenerationEU, partnership on “Telecommunications of the Future” (PE00000001 - program "RESTART”) and the Swedish Research Council (Grant 2022-03007 and 2024-04390).}
}

\maketitle
\thispagestyle{empty}

\begin{abstract}
High-mobility communications, which are crucial for next-generation wireless systems, cause the \gls{OFDM} waveform to suffer from strong \gls{ICI} due to the Doppler effect. In this work, we propose a novel receiver architecture for \gls{OFDM} that leverages the angular domain to separate multipaths. A block-type pilot is sent to estimate \glspl{DoA}, propagation delays, and channel gains of the multipaths. Subsequently, a \gls{DD} approach is employed to estimate and iteratively refine the Dopplers. Two different approaches are investigated to provide initial Doppler estimates: an \gls{EVM}-based method and a \gls{DL}-based method. Simulation results reveal that the \gls{DL}-based approach allows for constant \gls{BER} performance up to the maximum 6G speed of 1000 km/h.
\end{abstract}


\begin{comment}
    Channel estimation in OFDM systems is typically performed by sending pilot symbols over subcarriers to obtain the CFR. In particular, two different types of of pilot arrangements can be adopted: 
\begin{itemize}
    \item \textit{Block-type}: one OFDM symbol is fulfilled using pilot symbols and this OFDM block is used at the receiver to estimate the CFR. This configuration is usually adopted in static multipath channels where the CFR is almost constant during the entire transmission
    \item \textit{Comb-type}: a subset of the subcarriers is used to send pilot symbols during each OFDM block. This subset of subcarriers is used to obtain the CFR during each block by interpolating the response to obtain the entire CFR response over all the subcarriers. This approach is usually preferred in doubly-dispersive channels, where mobility makes the CFR change over time.

    In \cite{Colieri2002} different channel estimation schemes for OFDM are investigated and it is shown that comb-type pilot based channel estimation with low-pass
interpolation performs the best among all the considered channel estimation
algorithms as the comb-type pilot arrangement allows the tracking of time-varying channels and low-pass interpolation minimizes mean squared error.
\end{itemize}
\end{comment}

\glsresetall

\section{Introduction}
{T}{oward} the implementation of next-generation wireless systems, novel use cases are emerging, including \gls{HSR}, \gls{V2X}, \gls{UAV} operations, and autonomous vehicle scenarios. In these 6G high-mobility use cases, speeds of up to $1000$ km/h can be encountered \cite{Giordani2020,Tataria2021}.
The well-known \gls{OFDM} waveform, although suited for combating \gls{ISI} effects, suffers from \gls{ICI} caused by the Doppler effect and experiences performance degradation as mobility increases \cite{Wang2006,Hadani2017,Mostofi2005}. More recently, several waveforms have been introduced to overcome these \gls{OFDM} limitations \cite{Zhou2024}. Specifically, \gls{OTFS} modulation was introduced in \cite{Hadani2017}, leveraging communication in the delay-Doppler domain.
Typically, in \gls{OFDM} systems, channel estimation is performed by sending pilot sequences to estimate the \gls{CFR}. Conversely, in \gls{OTFS} systems, channel estimation is usually modeled as a parameter estimation problem due to the sparse nature of the channel in the delay-Doppler domain. Furthermore, channel parameters (channel gains, propagation delays, Doppler shifts and angles) typically exhibit longer coherence times compared to the \gls{CFR} \cite{Shaw2023}. This property is exploited in \cite{Shaw2023,Han2015}, where channel estimation in \gls{OFDM} systems is formulated as a parameter estimation problem to leverage the longer coherence time of delay-Doppler coordinates and reduce the number of parameters to be estimated. In \cite{Zhang2011,Ge2019,Feng2025,Guo2017}, angle-domain separability is exploited to compensate for Doppler shifts separately by transforming the multipath channel into multiple parallel single-Doppler channels. Solutions based on angle-domain processing for both sparse channels \cite{Zhang2011} and rich scattering environments \cite{Ge2019,Feng2025,Guo2017} have been investigated. However, in these studies, only the Doppler shift induced by the receiver mobility has been considered. Therefore, Doppler estimation is simply reduced to the estimation of the maximum Doppler shift \cite{Ge2019,Feng2025,Guo2017}. The single Doppler shifts affecting each path are then recovered and compensated based on the \gls{DoA} \cite{Zhang2011,Ge2019,Feng2025,Guo2017}. Moreover, performance investigation across the entire range of 6G speeds has not been done. Thus, the following question remains unanswered: \textit{is it possible to adopt angle-domain based multi-antenna \gls{OFDM} receivers for 6G high-mobility communications with multiple mobile scatterers?}
 
In order to answer this question, in this work, a \gls{DL}-empowered multi-antenna \gls{OFDM} receiver is developed by relaxing the assumptions made in \cite{Zhang2011,Ge2019,Feng2025,Guo2017}. Both parameter-based channel estimation \cite{Shaw2023,Han2015} and multipath separation in the angular domain \cite{Zhang2011} are exploited.
The contributions of this work can be summarized as follows:
\begin{itemize}
    \item \textbf{Multi-antenna \gls{OFDM} receiver with \gls{DD}-based multiple Doppler estimation}: A \gls{DoA}-aided \gls{OFDM} receiver using a block-type pilot \cite{Ge2019} is developed. Multipaths are separated in the angular domain and processed path-wise. Furthermore, a \gls{DD} approach, leveraging hard decisions as additional pilots, is used to estimate and refine multiple Dopplers by tracking the phase of time-varying channel gains.
    \item \textbf{Enhancement of initial Doppler estimate via \gls{DL}}: The zero Doppler initialization assumed in the \gls{DD} approach sets a maximum modulation-dependent speed that the receiver can handle. In order to overcome this limitation, a \gls{DL}-based approach is introduced to provide accurate initial Doppler estimates.
    \item \textbf{Mobility-resilience investigation for 6G deployments}: Simulations are carried out to investigate the robustness of the proposed \gls{DoA}-aided \gls{OFDM} receiver against mobility. Results show that the \gls{DL}-based approach allows performance to be reinforced against high mobility, achieving an almost constant \gls{BER} up to the maximum 6G speed of $1000$ km/h \cite{Giordani2020,Tataria2021}.
\end{itemize}

\begin{figure*}[t]
\centering
    \includegraphics[width=\textwidth]{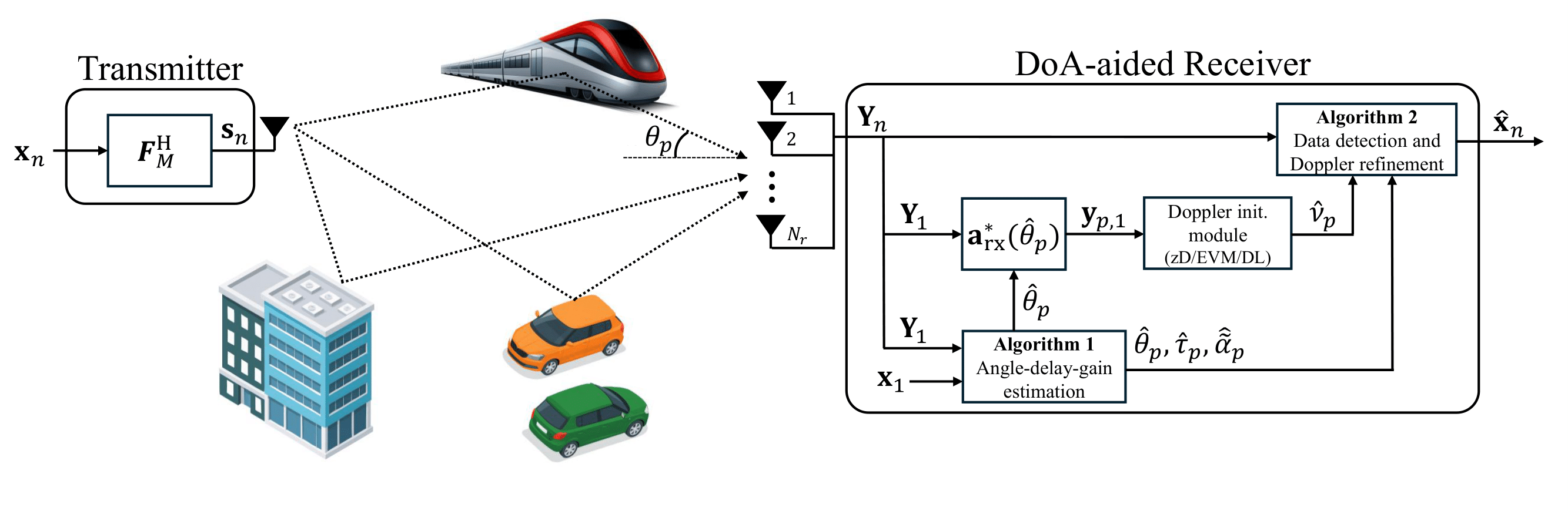}
    \caption{The OFDM system including a single-antenna transmitter, multiple reflectors (either static or mobile) and the proposed multi-antenna DoA-aided receiver containing a Doppler initialization module for coarse Doppler estimation from the received block-type pilot (zero Doppler, error vector magnitude or deep learning).}\label{fig:OFDMsystem} \vspace{-6mm}
\end{figure*}

\section{System Model}\label{sec:SysModel}
A scenario including a single-antenna \gls{TX} and a multi-antenna \gls{RX} equipped with a \gls{ULA} with $N_r$ receiving antennas is considered.
The system works at carrier frequency $f_c$. The antenna spacing at the \gls{RX} is $d=\lambda/2$, where $\lambda=c/f_c$ is the wavelength and $c$ denotes the speed of light.

\subsection{Single-Antenna OFDM Transmitter}
The \gls{TX} sends a block-type pilot to the \gls{RX} for channel estimation, followed by data frames using \gls{OFDM} signals. The \gls{OFDM} symbol is composed of $M$ subcarriers with spacing $\Delta f=1/T$, where $T$ is the symbol duration. Thus, the signal bandwidth is $B=M\Delta f$. As in classical \gls{OFDM} transmission, the transmit \gls{OFDM} symbol is preceded by a \gls{CP} with duration $T_{CP}>\sigma_\tau$, where $\sigma_\tau$ is the channel delay spread. Hence, the overall symbol duration is $T'=T+T_{CP}$. The delay and Doppler resolutions are $\Delta\tau=1/B=T/M$ and $\Delta\nu=\Delta f$, respectively. Moreover, the \gls{TX} sends $N$ \gls{OFDM} symbols within a geometric coherence time of the channel. Within this time, channel parameters (channel gains, propagation delays, Dopplers, and angles) are assumed to be constant \cite{Viterbo2022}. The \gls{TX} sends a block-type pilot followed by $N-1$ \gls{OFDM} data symbols.
The \gls{OFDM} modulator arranges symbols, taken from alphabet $\mathcal{C}$, in the frequency domain over the $M$ subcarriers. Hence, the symbol vector $\mathbf{x}_n\in\mathbb{C}^{M}$ is transmitted during the $n$-th symbol duration. The information symbols are normalized such that $\mathbb{E}\big[\big|[\mathbf{x}_n]_m\big|^2\big]=1$. The transmit signal is given as \cite{Keskin2021}
\begin{equation}
s_n(t)=\sqrt{\frac{P_T}{M}}\sum_{m=0}^{M-1}[\mathbf{x}_n]_me^{j2\pi m\Delta ft}\Pi\Big(\frac{t-nT'}{T'}\Big),
\end{equation}
where $\Pi(t)$ is a rectangular pulse that takes the value $1$ for $t\in[0,1]$ and $0$ otherwise, and $P_T$ is the average transmit power.

\begin{comment}
In particular, during the $n$-th symbol time the \gls{TX} sends
\begin{align} \label{eq:UEofdmsymbols}
\mathbf{x}_n=
    \begin{cases}
    \text{pilot} & ~n=1, \\
    \text{data}\in\mathcal{C} & ~n=2,3,...,N,
    \end{cases}
\end{align}
where $\mathcal{C}$ denotes the modulation alphabet (e.g. \gls{QAM}).
\end{comment}

\subsection{Observation Model at the Multi-Antenna Receiver}
Let's consider a transmission over a wireless channel made of $P$ propagation paths each with delay $\tau_p$, Doppler shift $\nu_p$, channel gain $\alpha_p$ and \gls{DoA} $\theta_p$. The channel response of the \gls{SIMO} channel is given as
\begin{equation}
\mathbf{h}(t,\tau)=\sum_{p=1}^P\alpha_pe^{j2\pi\nu_p t}\delta(\tau-\tau_p)\mathbf{a}_{\text{rx}}(\theta_p)\in\mathbb{C}^{N_r},
\end{equation}
where $\mathbf{a}_{\text{rx}}(\theta)=\big[1 \ \ e^{j\frac{2\pi}{\lambda}d\sin(\theta)} \ \ \dots \ \ e^{j\frac{2\pi}{\lambda}d(N_r-1)\sin(\theta)}\big]^T$ is the steering vector of the \gls{ULA} at the \gls{RX}.
During the $n$-th symbol time, the received signal at the $N_r$ antennas is obtained as
\begin{equation}
\begin{split}\label{eq:RxMultipathCP}
&\mathbf{y}_n^{\text{CP}}(t)=\int \mathbf{h}(t,\tau)s_n^{\text{CP}}(t-\tau)\text{d}\tau+\mathbf{n}(t)
\\
=&\sum_{p=1}^P\alpha_ps_n^{\text{CP}}(t-\tau_p)e^{j2\pi\nu_p t}\mathbf{a}_{\text{rx}}(\theta_p)+\mathbf{n}(t)\in\mathbb{C}^{N_r},
\end{split}
\end{equation}
where $\mathbf{n}(t)$ is \gls{AWGN} with a one-sided \gls{PSD} $N_0$, and $s_n^{\text{CP}}(t)$ denotes the transmit signal with the \gls{CP}.
After sampling at the symbol rate and \gls{CP} removal, the received time-spatial observations $\mathbf{Y}_n\in\mathbb{C}^{M\times N_r}$ during the $n$-th symbol time are obtained as \cite{Keskin2021,Keskin2024}
\begin{equation}\begin{split} \label{eq:timespatialobs}
\mathbf{Y}_n&=\sum_{p=1}^P\alpha_pe^{j2\pi\nu_p t_n}\mathbf{C}(\nu_p)\mathbf{F}_{M}^{\mathsf{H}}\mathbf{B}(\tau_p)\mathbf{F}_{M}\mathbf{s}_n\mathbf{a}_{\text{rx}}^\top(\theta_p)+\mathbf{N}
\\
=&\sqrt{P_T}\sum_{p=1}^P\tilde{\alpha}_{p,n}\big[\mathbf{F}_{M}^{\mathsf{H}}\big(\mathbf{x}_n\odot\mathbf{b}(\tau_p)\big)\odot\mathbf{c}(\nu_p)\big]\mathbf{a}_{\text{rx}}^\top(\theta_p)+\mathbf{N},
\end{split}\end{equation}
where
%\begin{itemize}
 %   \item 
 $\mathbf{s}_n=\sqrt{P_T}\mathbf{F}^{\mathsf{H}}_M\mathbf{x}_n\in\mathbb{C}^{M}$ is the transmit signal vector where $[\mathbf{F}_{M}]_{m,q}=\frac{1}{\sqrt{M}}e^{-j2\pi\frac{mq}{M}}$ denotes the $M$-point \gls{DFT} matrix. Moreover, $\mathbb{E}[\|\mathbf{s}_n\|^2]=MP_T$;
     $\mathbf{N}\in\mathbb{C}^{M\times N_r}$ is the \gls{AWGN} matrix and $\text{vec}(\mathbf{N})\sim\mathcal{CN}(\mathbf{0},\sigma^2\mathbf{I}_{MN_r})$. Noise variance is given as $\sigma^2=N_0B$ and $\mathbb{E}\big[\|\mathbf{N}\|_F^2\big]=MN_r\sigma^2$;
     $\mathbf{B}(\tau)=\text{diag}(\mathbf{b}(\tau))$, where $\mathbf{b}(\tau)=\big[e^{-j2\pi q\tau\Delta f}\big]_{q=0}^{M-1}$;  
     $\mathbf{C}(\nu)=\text{diag}(\mathbf{c}(\nu))$, where $\mathbf{c}(\nu)=\big[e^{j2\pi q\nu\Delta\tau}\big]_{q=0}^{M-1}$ captures fast-time effects caused by Doppler (i.e. \gls{ICI});  $\tilde{\alpha}_{p,n}=\alpha_pe^{j2\pi\nu_p t_n}$ is the Doppler-induced time varying channel gain that captures slow-time variations where $t_n=nT_{\text{CP}}+(n-1)T$ is the time at which the transmission of the $n$-th \gls{OFDM} symbol starts.  
%\end{itemize}
The \gls{SNR} is obtained as
$\text{SNR}={\|\boldsymbol{\alpha}\|^2P_T}/{(N_0B)}$ where $\boldsymbol{\alpha}=[\alpha_1, \alpha_2,...,\alpha_P]^\top$ is the channel gains vector.

\section{Proposed DoA-Aided OFDM Receiver}
In this section the architecture of the proposed \gls{DoA}-aided receiver is presented. First, multipath separation in the angle-domain is discussed; afterwards, delay compensation, \gls{ICI} cancellation and channel parameter estimation with \gls{DD} Doppler estimation are presented.
\subsection{Path Separation via Angle-domain Matched Filter}
Assuming that the \gls{RX} is equipped with a large \gls{ULA}, different directions are approximately orthogonal \cite{Ge2019}; i.e. $\mathbf{a}_{\text{rx}}^\top(\theta_1)\mathbf{a}_{\text{rx}}^*(\theta_2)\approx 0$ if $\theta_1\neq\theta_2$. Consequently, a \gls{MF} can be applied to separate multipaths by computing 
\begin{equation}\begin{split}\label{eq:AngleDomainMF}
\mathbf{y}_{p,n}=\frac{\mathbf{Y}_n\mathbf{a}_{\text{rx}}^*(\theta_p)}{N_r}=\sqrt{P_T} \tilde{\alpha}_{p,n}\big[\mathbf{F}_{M}^{\mathsf{H}}\big(\mathbf{x}_n\odot\mathbf{b}(\tau_p)\big)\odot\mathbf{c}(\nu_p)\big]+&
\\
\underbrace{\sqrt{P_T}\sum_{\substack{i=1 \\ i\neq p}}^P\tilde{\alpha}_{i,n} \mathbf{a}_{\text{rx}}^\top(\theta_i)\mathbf{a}_{\text{rx}}^*(\theta_p)\big[\mathbf{F}_{M}^{\mathsf{H}}\big(\mathbf{x}_n\odot\mathbf{b}(\tau_i)\big)\odot\mathbf{c}(\nu_i)\big]}_{\text{IPI}}+\mathbf{n}_p&
\\
\approx\sqrt{P_T}\tilde{\alpha}_{p,n}\big[\mathbf{F}_{M}^{\mathsf{H}}\big(\mathbf{x}_n\odot\mathbf{b}(\tau_p)\big)\odot\mathbf{c}(\nu_p)\big]+\mathbf{n}_p\in\mathbb{C}^{M},&
\end{split}\end{equation}
where the second term is known as \gls{IPI} and approaches zero as $N_r\rightarrow\infty$.
Moreover, $\mathbf{n}_p=\mathbf{N}\mathbf{a}_{\text{rx}}^*(\theta_p)/N_r$ is noise with variance $\sigma_p^2=\mathbb{E}[\|\mathbf{n}_p\|^2]/M$. Since $\mathbb{E}[\mathbf{N}^{\mathsf{H}}\mathbf{N}]=M\sigma^2\mathbf{I}_{N_r}$ and $\|\mathbf{a}_{\text{rx}}(\theta_p)\|^2=N_r$, the noise variance after beamforming is $\sigma_p^2=\sigma^2/N_r$.

\subsection{Data Detection: Delay/ICI Compensation}\label{ISIICIcompensation}
In this section, data detection is discussed assuming knowledge of channel parameters. The architecture of the \gls{DoA}-aided receiver performing both estimation and data detection is discussed in the next section.
\subsubsection{\gls{ICI} Cancellation via Time-Domain Single-tap \gls{MF}}
Given the impaired observation $\mathbf{y}_{p,n}$, the effect of the Doppler shift $\nu_p$ can be compensated for by computing\footnote{Considering a single-path impaired signal, \gls{ICI} can be easily compensated for by using $\mathbf{c}(\nu)\odot\mathbf{c}^*(\nu)=\mathbf{1}$. The same holds also for the delay shift.}
\begin{equation} \label{eq:ICIcancelaltionMF}
\mathbf{y}_{p,n}^{\text{ICI}}=\mathbf{y}_{p,n}\odot\mathbf{c}^*(\hat{\nu}_p)\approx\sqrt{P_T}\tilde{\alpha}_{p,n}\mathbf{F}_{M}^{\mathsf{H}}\big(\mathbf{x}_n\odot\mathbf{b}(\tau_p)\big)+\mathbf{n}_p.
\end{equation}

\subsubsection{Delay compensation via Frequency-Domain Single-tap \gls{MF}}
Given the \gls{ICI}-compensated observation $\mathbf{y}_{p,n}^{\text{ICI}}$, the effect of the propagation delay $\tau_p$ can be compensated for by computing the frequency domain \gls{ICI}-compensated symbols
\begin{equation} \label{eq:TDtoFD}
\mathbf{x}_{p,n}^{\text{ICI}}=\mathbf{F}_M\mathbf{y}_{p,n}^{\text{ICI}}\approx\sqrt{P_T}\tilde{\alpha}_{p,n}\big(\mathbf{x}_n\odot\mathbf{b}(\tau_p)\big)+\mathbf{n}_p,
\end{equation}
and applying a single-tap \gls{MF} as
\begin{equation} \label{eq:ISIcancelaltionMF}
\hat{\mathbf{x}}_{p,n}=\mathbf{x}_{p,n}^{\text{ICI}}\odot\mathbf{b}^*(\hat{\tau}_p)\approx\sqrt{P_T}\tilde{\alpha}_{p,n}\mathbf{x}_n+\mathbf{n}_p.
\end{equation}

\subsubsection{Maximum ratio combining}
Following the previous steps, $P$ estimates of the information symbols are obtained ($\hat{\mathbf{x}}_{p,n}$). In order to maximize \gls{SNR}, \gls{MRC} can be used to obtain the estimated symbols as
%\begin{equation} \label{eq:MRC}
$\hat{\mathbf{x}}_n=\sum_{p=1}^P\tilde{\alpha}_{p,n}^*\hat{\mathbf{x}}_{p,n}\approx\sqrt{P_T}\|\boldsymbol{\alpha}_n\|^2\mathbf{x}_n+\sum_{p=1}^P\tilde{\alpha}_{p,n}^*\mathbf{n}_p$,
%\end{equation}
where $\boldsymbol{\alpha}_n=[\tilde\alpha_{1,n},\tilde\alpha_{2,n},...,\tilde\alpha_{P,n}]^\top$.
In conclusion, assuming accurate parameter estimation and negligible \gls{IPI} due to large number of antenna elements, the proposed \gls{DoA}-aided receiver transforms the doubly-dispersive channel into an equivalent channel affected by additive noise.

\begin{comment}
\begin{figure}[t]
\centering
    \includegraphics[width=0.99\columnwidth]{Results/AngleEstimation.png}
    \caption{The received power is shown as a function of the DOA. The multipaths are detected in the angular domain by looking for peaks exceeding a predefined threshold.}\label{fig:AngleEstim}
\end{figure}
\end{comment}

\begin{algorithm}[t]
\caption{Proposed angle-delay-gain estimation}\label{DOAestimAlg}
\KwIn{ $\mathbf{Y}_1$, $\mathbf{x}_1$}
\textbf{Path detection and DoA estimation:} \\
Run CFAR to $\mathcal{P}(\theta)=\|\mathbf{Y}_1\mathbf{a}_{\text{rx}}^*(\theta)\|^2$ and obtain estimated DoAs $\{\hat{\theta}_p\}_{p=1}^{\hat{P}}$\\
\textbf{Delay-gain estimation:} \\
\For{$p = 1$ \KwTo $\hat{P}$}{
  $\mathbf{y}_{p,1}=\frac{\mathbf{Y}_1\mathbf{a}_{\text{rx}}^*(\hat{\theta}_p)}{N_r}$ \\  $\hat{\tau}_p=\arg\max_{\tau}\big|\mathbf{\tilde{b}}^{\mathsf{H}}(\tau)\mathbf{y}_{p,1}\big|$\\
 $\hat{\tilde{\alpha}}_{p,1}=\frac{\mathbf{\tilde{b}}(\hat{\tau}_p)^{\mathsf{H}}\mathbf{y}_{p,1}}{M\sqrt{P_T}}$
}
\KwOut{$\{\hat{\theta}_p,\hat{\tau}_p,\hat{\tilde{\alpha}}_{p,1}\}_{p=1}^{\hat{P}}$}
\end{algorithm}

\subsection{Channel Parameter Estimation}\label{sec_estim&detect}
Given the \gls{OFDM} system model in Section \ref{sec:SysModel}, the following estimation algorithm is proposed. A more detailed description is provided in Algorithm \ref{DOAestimAlg} and Algorithm \ref{DOAaidedRXAlg}.

\subsubsection{Angle Estimation}
The estimation of the \gls{DoA} for each path is carried out by computing the received power as a function of the direction $\theta$ as $\mathcal{P}(\theta)=\|\mathbf{Y}_1\mathbf{a}_{rx}^*(\theta)\|^2$. Thus, the \glspl{DoA} can be estimated by searching for peaks in the angular spectrum by running a \gls{CFAR} detector to set an adaptive threshold.

\subsubsection{Angle-domain MF}
A \gls{MF} is applied to separate the detected multipaths, as shown in \eqref{eq:AngleDomainMF}. Therefore, the received block-type pilot is given as
$\mathbf{y}_{p,1}=\sqrt{P_T}\tilde{\alpha}_{p,1}\mathbf{\tilde{b}}(\tau_p)\odot\mathbf{c}(\nu_p)+\mathbf{n}_p$, where $\mathbf{\tilde{b}}(\tau)=\mathbf{F}_{M}^{\mathsf{H}}\big(\mathbf{x}_1\odot\mathbf{b}(\tau)\big)$ is the time-domain delay term.

\subsubsection{Delay Estimation}
If the coherence time\footnote{The coherence time is given as $(\Delta t)_{c}\approx\frac{1}{\sigma_{\nu}}$, where $\sigma_{\nu}=\frac{f_c}{c}v_{\max}$ is the Doppler spread and $v_{\max}$ is the maximum speed.} $(\Delta t)_{c}$ becomes comparable with $T'$, the Doppler-induced \gls{ICI} can be neglected to perform delay estimation. This approximation is equivalent to considering a constant channel within an \gls{OFDM} symbol transmission, and therefore, there is no \gls{ICI} \cite{Shaw2023}. However, the Doppler must be estimated since it makes the channel response change over time, and \gls{ICI} should be compensated to avoid performance degradation in data detection. Assuming that this condition holds due to poor Doppler resolution, the MF output can be approximated by $\mathbf{y}_{p,1}\approx\sqrt{P_T}\tilde{\alpha}_{p,1}\mathbf{\tilde{b}}(\tau_p)+\mathbf{n}_p$.
Finally, the delay can be estimated as follows:
$\hat{\tau}_p=\arg\max_{\tau}\big|\mathbf{\tilde{b}}^{\mathsf{H}}(\tau)\mathbf{y}_{p,1}\big|$.

\subsubsection{Gain Estimation}
Once a delay-angle pair is obtained, the estimation of the channel gain associated with the the $p$-th path can be done using the \gls{LS} estimator as $\hat{\tilde{\alpha}}_{p,1}=(\mathbf{\tilde{b}}(\hat{\tau}_p)^{\mathsf{H}}\mathbf{y}_{p,1})/(M\sqrt{P_T})$.

\subsubsection{Decision-Directed Doppler Estimation via Gain Tracking}
A \gls{DD} approach on the incoming \gls{OFDM} data symbols is used to obtain an estimate of the Doppler by increasing the observation time.
In particular, a sliding window of length $2<K<N-1$\footnote{The parameter $K$ should be selected to avoid wrapping issues. In particular, given the channel Doppler spread $\sigma_\nu$, $K$ should satisfy $2\pi\sigma_{\nu}(K-1)T'<\pi$. Equivalently, $K<1+\frac{1}{2\sigma_\nu T'}$.} is considered to estimate Dopplers. A slicer is adopted to obtain hard decisions (denoted $\hat{\mathbf{x}}_n^\dagger=\mathcal{C}(\hat{\mathbf{x}}_n)$, where $\mathcal{C}(\cdot)$ is the hard decision according to alphabet $\mathcal{C}$), and decoded data are used as additional pilots to estimate time-varying channel gains through \gls{LS}.
Once channel gains are obtained, the Doppler can be estimated by noting that $\phi_p=\angle\hat{\tilde{\alpha}}_{p,n+K-1}\hat{\tilde{\alpha}}_{p,n}^*\approx2\pi\nu_p(t_{n+K-1}-t_n)=2\pi\nu_p (K-1)T'$.
Thus, $\hat{\nu}_p=\phi_p/({2\pi(K-1)T'})$.

\begin{algorithm}[t]
\caption{Proposed DoA-aided receiver}\label{DOAaidedRXAlg}
\KwIn{ $\mathbf{Y}_n$, $\mathbf{x}_1$, $K$, \\ \ \ \ \ \ \ \ \ \
$\{\hat{\theta}_p,\hat{\tau}_p,\hat{\tilde{\alpha}}_{p,1}\}_{p=1}^{\hat{P}}$ obtained using Algorithm \ref{DOAestimAlg}}

\textbf{Initialize}: $\hat{\nu}_p=0$ (zero-Doppler)

\textbf{Joint data detection and Doppler estimation:} \\
\For{$n = 2$ \KwTo $N-K+1$}{
  \uIf{$n < N - K + 1$}{
    \For{$k = 1$ \KwTo $K$}{
      \For{$p = 1$ \KwTo $\hat{P}$}{
        $\mathbf{y}_{p,n+k-1} = \frac{\mathbf{Y}_{n+k-1}\mathbf{a}_{\text{rx}}^*(\hat{\theta}_p)}{N_r}$ \\
        $\mathbf{y}_{p,n+k-1}^{\text{ICI}} = \mathbf{y}_{p,n+k-1} \odot \mathbf{c}^*(\hat{\nu}_p)$ \\
        $\hat{\mathbf{x}}_{p,n+k-1} = \mathbf{F}_M\mathbf{y}_{p,n+k-1}^{\text{ICI}}\odot \mathbf{b}^*(\hat{\tau}_p)$ \\
        $\hat{\tilde{\alpha}}_{p,n+k-1}=\hat{\tilde{\alpha}}_{p,n+k-2}  \ e^{j 2\pi \hat{\nu}_p T'}$
      }
      $\hat{\mathbf{x}}_{n+k-1}=\sum_{p=1}^{\hat{P}}\hat{\tilde{\alpha}}_{p,n+k-1}^*\hat{\mathbf{x}}_{p,n+k-1}$ \\
      $\hat{\mathbf{x}}_{n+k-1}^\dagger=\mathcal{C}(\hat{\mathbf{x}}_{n+k-1})$\\
      \For{$p = 1$ \KwTo $\hat{P}$}{
        $\hat{\tilde{\alpha}}_{p,n+k-1}=\frac{\big[\mathbf{F}^{\mathsf{H}}_M(\hat{\mathbf{x}}_{n+k-1}^\dagger\odot\mathbf{b}(\hat{\tau}_p))\odot\mathbf{c}(\hat{\nu}_p)\big]^{\mathsf{H}}\mathbf{y}_{p,n+k-1}}{M\sqrt{P_T}}$
      }
    }

    $\hat{\nu}_p = \dfrac{\angle\hat{\tilde{\alpha}}_{p,n+K-1}\hat{\tilde{\alpha}}_{p,n}^*}{2\pi (K-1) T'}$

    \For{$p = 1$ \KwTo $\hat{P}$}{
        $\hat{\mathbf{x}}_{p,n} = \mathbf{F}_M(\mathbf{y}_{p,n} \odot \mathbf{c}^*(\hat{\nu}_p))\odot \mathbf{b}^*(\hat{\tau}_p)$
    }
      $\hat{\mathbf{x}}_{n}=\sum_{p=1}^{\hat{P}}\hat{\tilde{\alpha}}_{p,n}^*\hat{\mathbf{x}}_{p,n}$ \\
      $\hat{\mathbf{x}}_{n}^\dagger=\mathcal{C}(\hat{\mathbf{x}}_{n})$\\
    }
  }
  \Else{
    \For{$k = 1$ \KwTo $K$}{
      \For{$p = 1$ \KwTo $\hat{P}$}{
      $\mathbf{y}_{p,n+k-1} = \frac{\mathbf{Y}_{n+k-1}\mathbf{a}_{\text{rx}}^*(\hat{\theta}_p)}{N_r}$ \\
        $\mathbf{y}_{p,n+k-1}^{\text{ICI}} = \mathbf{y}_{p,n+k-1} \odot\mathbf{c}^*(\hat{\nu}_p)$ \\
        $\hat{\mathbf{x}}_{p,n+k-1} = \mathbf{F}_M\mathbf{y}_{p,n+k-1}^{\text{ICI}}\odot \mathbf{b}^*(\hat{\tau}_p)$ \\
        $\hat{\tilde{\alpha}}_{p,n+k-1}=\hat{\tilde{\alpha}}_{p,n-1} \ e^{j 2\pi \hat{\nu}_p k T'}$
      }
     $\hat{\mathbf{x}}_{n+k-1}=\sum_{p=1}^{\hat{P}}\hat{\tilde{\alpha}}_{p,n+k-1}^*\hat{\mathbf{x}}_{p,n+k-1}$ \\
      $\hat{\mathbf{x}}_{n+k-1}^\dagger=\mathcal{C}(\hat{\mathbf{x}}_{n+k-1})$
    }
  }

\KwOut{$\hat{\mathbf{x}}_{n}^\dagger$}
\end{algorithm}

\subsection{Enhancing Initial Doppler Estimate}

Due to the \gls{zD} initialization in Algorithm \ref{DOAaidedRXAlg}, the performance of the proposed approach is limited up to a maximum Doppler of $\nu_{\max}=\frac{1}{8T'}$ \footnote{The maximum Doppler supported by the proposed approach can be computed by noting that the maximum large-scale Doppler-induced phase rotation can be at most $\pi/4$ when using 4-\gls{QAM} signaling. Therefore, $2\pi\nu_{\max}T'<\frac{\pi}{4}$.} for 4-\gls{QAM}, primarily because of the rapidly varying phase of the channel gain $\tilde{\alpha}_{p,n}$. Consequently, the hard decisions $\hat{\mathbf{x}}_n^\dagger$ are subject to decision errors that stem solely from the uncompensated Doppler-induced phase rotation resulting from the \gls{zD} initialization assumed in Section \ref{sec_estim&detect}. In order to enhance performance at higher speeds without increasing $\Delta f$ (and thereby deteriorating spectral efficiency), an initial guess of the Doppler must be provided. This prevents error propagation by enabling the \gls{RX} to track rapid phase rotations in subsequent \gls{OFDM} symbols. The following approaches are investigated:

\subsubsection{EVM Minimization}
The Doppler effect causes \gls{ICI} in the signal within each branch, which, in turn, increases the \gls{EVM}. Therefore, an initial guess of the Doppler can be obtained by searching for a coarse Doppler shift that minimizes the \gls{EVM}. Thus $\hat{\nu}_{p}=\arg\min_\nu \text{EVM}_p(\nu)$,
where
\begin{equation}
\text{EVM}_p(\nu)=\frac{1}{M}\Bigg\|\frac{\mathbf{F}_M\big[\mathbf{y}_{p,1}\odot\mathbf{c}^*(\nu)\big]\odot\mathbf{b}^*(\hat{\tau}_p)}{\hat{\tilde{\alpha}}_{p,1}\sqrt{P_T}}-\mathbf{x}_1\Bigg\|^2 .
\end{equation}

\subsubsection{DL-based Approach}
A \gls{DL} architecture, which performs a regression task, is used to predict the Doppler shift from the received block-type pilot. In particular, a \gls{FNN} receives $\mathbf{y}_{p,1}/(\hat{\tilde{\alpha}}_{p,1}\sqrt{P_T})$ as input and provides a coarse estimate of the Doppler affecting the $p$-th path ($\hat\nu$) as output. The objective function for the optimization problem is the \gls{MSE} $\mathcal{L}(\hat{\nu},\nu)=|\hat{\nu}-\nu|^2$.

\begin{figure*}[t] 
    \centering  
    \begin{subfigure}[t]{0.31\textwidth}
        \centering
        \resizebox{0.98\columnwidth}{!}{
        \input{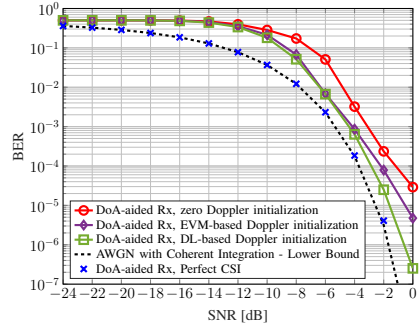}
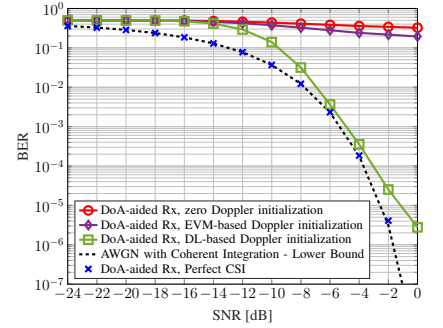
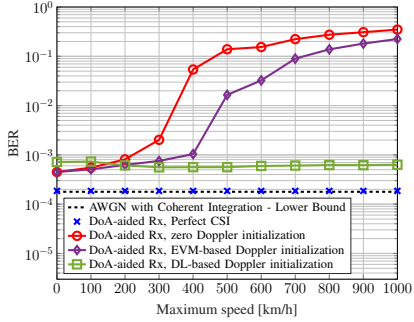
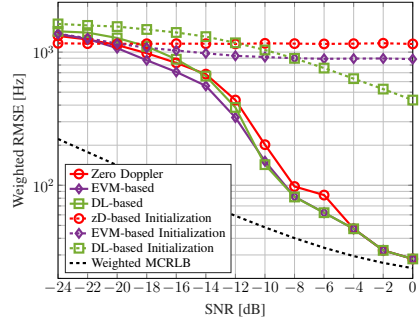
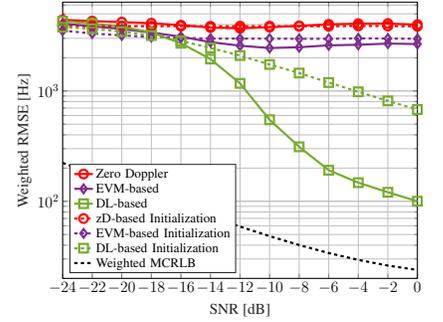}
    \caption{DL-based Doppler predictions are shown against true values. Both are normalized with respect to the maximum Doppler shift used during training.}\label{fig:testDoppler}
    \end{subfigure}
    \hfill   
\begin{subfigure}[t]{0.31\textwidth}
        \centering
        \resizebox{0.98\columnwidth}{!}{
        % Questo file è stato creato da matlab2tikz.
%
% Gli ultimi aggiornamenti possono essere recuperati da
% dove puoi anche fare suggerimenti e votare matlab2tikz.
%
\definecolor{mycolor1}{rgb}{0.49400,0.18400,0.55600}%
\definecolor{mycolor2}{rgb}{0.46600,0.67400,0.18800}%

\begin{tikzpicture}

\begin{axis}[%
width=4.521in,
height=3.566in,
at={(0.758in,0.481in)},
scale only axis,
xmin=-24,
xmax=0,
% SINTASSI CORRETTA per lo stile del font: racchiudi tutto tra graffe
xlabel style={font=\Large\color{white!15!black},yshift=-5pt},
xlabel={SNR [dB]},
ymode=log,
ymin=1e-07,
ymax=1,
yminorticks=true,
% SINTASSI CORRETTA per lo stile del font: racchiudi tutto tra graffe
ylabel style={font=\Large\color{white!15!black},yshift=7pt},
ylabel={BER},
axis background/.style={fill=white},
title style={font=\bfseries},
tick label style={font=\Large},
xmajorgrids,
ymajorgrids,
yminorgrids,
legend style={at={(0.02,0.02)}, anchor=south west, legend cell align=left, align=left, draw=white!15!black, font=\large}
]

\addplot [color=red, line width=2pt, mark=o, mark options={solid, red, mark size=4pt}]
  table[row sep=crcr]{%
-24	0.500004070060484\\
-22	0.499952394153226\\
-20	0.499936076108871\\
-18	0.499201839717742\\
-16	0.492863571068548\\
-14	0.468803011592742\\
-12	0.396216645665323\\
-10	0.284044682459677\\
-8	0.172891330645161\\
-6	0.0508929813508065\\
-4	0.00321354586693548\\
-2	0.000235496471774194\\
0	2.90826612903226e-05\\
};
\addlegendentry{DoA-aided Rx, zero Doppler initialization}

\addplot [color=mycolor1, line width=2pt, mark=diamond, mark options={solid, mycolor1, mark size=4pt}]
  table[row sep=crcr]{%
-24	0.50001033266129\\
-22	0.49994666078629\\
-20	0.49978107358871\\
-18	0.498022139616935\\
-16	0.487461391129032\\
-14	0.452817893145161\\
-12	0.354329964717742\\
-10	0.214973072076613\\
-8	0.0676400075604839\\
-6	0.0068488533266129\\
-4	0.00086281502016129\\
-2	7.87676411290323e-05\\
0	4.75050403225806e-06\\
};
\addlegendentry{DoA-aided Rx, EVM-based Doppler initialization}

\addplot [color=mycolor2, line width=2pt, mark=square, mark options={solid, mycolor2, mark size=4pt}]
  table[row sep=crcr]{%
-24	0.500006224798387\\
-22	0.499953641633064\\
-20	0.499741557459677\\
-18	0.497918623991935\\
-16	0.484767993951613\\
-14	0.43839314516129\\
-12	0.338100012600806\\
-10	0.180330330141129\\
-8	0.0514830519153226\\
-6	0.00669782006048387\\
-4	0.000637676411290323\\
-2	2.5e-05\\
0	2.52016129032258e-07\\
};
\addlegendentry{DoA-aided Rx, DL-based Doppler initialization}

\addplot [color=black, dashed, line width=2pt]
  table[row sep=crcr]{%
-24	0.36057158203125\\
-22	0.326584704589844\\
-20	0.285800219726562\\
-18	0.238177294921875\\
-16	0.184989880371094\\
-14	0.129540148925781\\
-12	0.0776760620117188\\
-10	0.03683203125\\
-8	0.01216796875\\
-6	0.00229388427734375\\
-4	0.0001792724609375\\
-2	3.16162109375e-06\\
0	2.4414e-9\\
};
\addlegendentry{AWGN with Coherent Integration - Lower Bound}

\addplot [color=blue, line width=2pt, only marks, mark=x, mark options={solid, blue, mark size=4pt}]
  table[row sep=crcr]{%
-24	0.36061767578125\\
-22	0.326612341308594\\
-20	0.285840698242187\\
-18	0.238232299804687\\
-16	0.185099938964844\\
-14	0.12954111328125\\
-12	0.0777566772460938\\
-10	0.0368931884765625\\
-8	0.012215087890625\\
-6	0.00232432861328125\\
-4	0.00018441162109375\\
-2	4.08935546875e-06\\
0	1.220703125e-08\\
};
\addlegendentry{DoA-aided Rx, Perfect CSI}

\end{axis}

\end{tikzpicture}%
}
    \caption{The BER is shown against SNR for a maximum UE speed of $300$ km/h. The proposed DoA-aided approach in doubly-dispersive channel is compared against AWGN with coherent integration.}\label{fig:bervssnr300}
    \end{subfigure}
    \hfill 
\begin{subfigure}[t]{0.31\textwidth}
        \centering
        \resizebox{0.98\columnwidth}{!}{
        % Questo file è stato creato da matlab2tikz.
%
% Gli ultimi aggiornamenti possono essere recuperati da
% dove puoi anche fare suggerimenti e votare matlab2tikz.
%
\definecolor{mycolor1}{rgb}{0.49400,0.18400,0.55600}%
\definecolor{mycolor2}{rgb}{0.46600,0.67400,0.18800}%

\begin{tikzpicture}

\begin{axis}[%
width=4.521in,
height=3.566in,
at={(0.758in,0.481in)},
scale only axis,
xmin=-24,
xmax=0,
% SINTASSI CORRETTA per lo stile del font: racchiudi tutto tra graffe
xlabel style={font=\Large\color{white!15!black},yshift=-5pt},
xlabel={SNR [dB]},
ymode=log,
ymin=1e-07,
ymax=1,
yminorticks=true,
% SINTASSI CORRETTA per lo stile del font: racchiudi tutto tra graffe
ylabel style={font=\Large\color{white!15!black},yshift=7pt},
ylabel={BER},
axis background/.style={fill=white},
title style={font=\bfseries},
tick label style={font=\Large},
xmajorgrids,
ymajorgrids,
yminorgrids,
legend style={at={(0.02,0.02)}, anchor=south west, legend cell align=left, align=left, draw=white!15!black, font=\large}
]

\addplot [color=red, line width=2pt, mark=o, mark options={solid, red, mark size=4pt}]
  table[row sep=crcr]{%
-24	0.49865\\
-22	0.497897\\
-20	0.495964\\
-18	0.490983\\
-16	0.492067\\
-14	0.480196\\
-12	0.461332\\
-10	0.43774\\
-8	0.410922\\
-6	0.38372\\
-4	0.358926\\
-2	0.341309\\
0	0.325301\\
};
\addlegendentry{DoA-aided Rx, zero Doppler initialization}

\addplot [color=mycolor1, line width=2pt, mark=diamond, mark options={solid, mycolor1, mark size=4pt}]
  table[row sep=crcr]{%
-24	0.49865\\
-22	0.497897\\
-20	0.495964\\
-18	0.490983\\
-16	0.483606\\
-14	0.459702\\
-12	0.421131\\
-10	0.374794\\
-8	0.325835\\
-6	0.279843\\
-4	0.240417\\
-2	0.216571\\
0	0.194553\\
};
\addlegendentry{DoA-aided Rx, EVM-based Doppler initialization}

\addplot [color=mycolor2, line width=2pt, mark=square, mark options={solid, mycolor2, mark size=4pt}]
  table[row sep=crcr]{%
-24	0.49865\\
-22	0.497897\\
-20	0.495964\\
-18	0.490983\\
-16	0.473943\\
-14	0.414046\\
-12	0.293069\\
-10	0.13981\\
-8	0.0316119\\
-6	0.00363535\\
-4	0.000354456\\
-2	2.53503e-05\\
0	2.78226e-06\\
};
\addlegendentry{DoA-aided Rx, DL-based Doppler initialization}

\addplot [color=black, dashed, line width=2pt]
  table[row sep=crcr]{%
-24	0.36057158203125\\
-22	0.326584704589844\\
-20	0.285800219726562\\
-18	0.238177294921875\\
-16	0.184989880371094\\
-14	0.129540148925781\\
-12	0.0776760620117188\\
-10	0.03683203125\\
-8	0.01216796875\\
-6	0.00229388427734375\\
-4	0.0001792724609375\\
-2	3.16162109375e-06\\
0	2.4414e-9\\
};
\addlegendentry{AWGN with Coherent Integration - Lower Bound}

\addplot [color=blue, line width=2pt, only marks, mark=x, mark options={solid, blue, mark size=4pt}]
  table[row sep=crcr]{%
-24	0.36061767578125\\
-22	0.326612341308594\\
-20	0.285840698242187\\
-18	0.238232299804687\\
-16	0.185099938964844\\
-14	0.12954111328125\\
-12	0.0777566772460938\\
-10	0.0368931884765625\\
-8	0.012215087890625\\
-6	0.00232432861328125\\
-4	0.00018441162109375\\
-2	4.08935546875e-06\\
0	1.220703125e-08\\
};
\addlegendentry{DoA-aided Rx, Perfect CSI}

\end{axis}

\end{tikzpicture}%
}
    \caption{The BER is shown against SNR for a maximum UE speed of $1000$ km/h. The proposed DoA-aided approach in doubly-dispersive channel is compared against AWGN with coherent integration.}\label{fig:bervssnr1000}
    \end{subfigure}

    \par
    \begin{subfigure}[t]{0.31\textwidth}
        \centering
        \resizebox{0.98\columnwidth}{!}{
        \definecolor{mycolor1}{rgb}{0.49400,0.18400,0.55600}%
\definecolor{mycolor2}{rgb}{0.46600,0.67400,0.18800}%

\begin{tikzpicture}

\begin{axis}[%
width=4.521in,
height=3.617in,
at={(0.758in,0.488in)},
scale only axis,
xmin=0,
xmax=1000,
xticklabel style={
    /pgf/number format/set thousands separator={},
    /pgf/number format/1000 sep={}, % Utile per compatibilità con versioni meno recenti
    font=\Large % Mantieni la dimensione del font che avevi impostato
},
% CORREZIONE QUI: "font=" deve precedere "\Large"
xlabel style={font=\Large\color{white!15!black},yshift=-5pt},
xlabel={Maximum speed [km/h]},
ymode=log,
ymin=3*1e-06,
ymax=1,
yminorticks=true,
% CORREZIONE QUI: "font=" deve precedere "\Large"
ylabel style={font=\Large\color{white!15!black},yshift=7pt},
ylabel={BER},
axis background/.style={fill=white},
% Questa riga era già corretta, la manteniamo
tick label style={font=\Large},
xmajorgrids,
ymajorgrids,
yminorgrids,
legend style={at={(0.02,0.02)}, anchor=south west, legend cell align=left, align=left, draw=white!15!black,  font=\large}
]
\addplot [color=black, dashed, line width=2pt]
  table[row sep=crcr]{%
0	0.000178241271077713\\
100	0.000178241271077713\\
200	0.000178241271077713\\
300	0.000178241271077713\\
400	0.000178241271077713\\
500	0.000178241271077713\\
600	0.000178241271077713\\
700	0.000178241271077713\\
800	0.000178241271077713\\
900	0.000178241271077713\\
1000	0.000178241271077713\\
};
\addlegendentry{AWGN with Coherent Integration - Lower Bound}

\addplot [color=blue, line width=2pt, only marks, mark=x, mark options={solid, blue, mark size=4pt}]
  table[row sep=crcr]{%
0	0.000185618470491202\\
100	0.000185618470491202\\
200	0.000185618470491202\\
300	0.000185618470491202\\
400	0.000185618470491202\\
500	0.000185618470491202\\
600	0.000185618470491202\\
700	0.000185618470491202\\
800	0.000185618470491202\\
900	0.000185618470491202\\
1000	0.000185618470491202\\
};
\addlegendentry{DoA-aided Rx, Perfect CSI}

\addplot [color=red, line width=2pt, mark=o, mark options={solid, red, mark size=4pt}]
  table[row sep=crcr]{%
0	0.000448210685483871\\
100	0.00055078125\\
200	0.000809286794354839\\
300	0.00201864919354839\\
400	0.0540598853326613\\
500	0.139017641129032\\
600	0.154055758568548\\
700	0.22291951234879\\
800	0.273786479334677\\
900	0.308505796370968\\
1000	0.348109847530242\\
};
\addlegendentry{DoA-aided Rx, zero Doppler initialization}

\addplot [color=mycolor1, line width=2pt, mark=diamond, mark options={solid, mycolor1,mark size=4pt}]
  table[row sep=crcr]{%
0	0.000448210685483871\\
100	0.000509198588709677\\
200	0.000624968497983871\\
300	0.000753181703629032\\
400	0.00103720388104839\\
500	0.0164680569556452\\
600	0.0323901209677419\\
700	0.0897579700100807\\
800	0.138342552923387\\
900	0.181009482106855\\
1000	0.224475396925403\\
};
\addlegendentry{DoA-aided Rx, EVM-based Doppler initialization}

\addplot [color=mycolor2, line width=2pt, mark=square, mark options={solid, mycolor2,mark size=4pt}]
  table[row sep=crcr]{%
0	0.000717899445564516\\
100	0.000731287802419355\\
200	0.000611485635080645\\
300	0.000559097782258064\\
400	0.000562752016129032\\
500	0.000563760080645161\\
600	0.000594821068548387\\
700	0.000607705393145161\\
800	0.000621660786290323\\
900	0.000619518649193548\\
1000	0.000631741431451613\\
};
\addlegendentry{DoA-aided Rx, DL-based Doppler initialization}

\end{axis}

\end{tikzpicture}%
}
    \caption{The BER is shown against maximum UE velocity with a fixed SNR$=-4$ dB. The proposed DoA-aided approach in doubly-dispersive channel is compared against AWGN with coherent integration.}\label{fig:bervsspeed}
    \end{subfigure}
    \hfill
    \begin{subfigure}[t]{0.31\textwidth}
        \centering
        \resizebox{0.98\columnwidth}{!}{
        \definecolor{mycolor1}{rgb}{0.49400,0.18400,0.55600}%
\definecolor{mycolor2}{rgb}{0.46600,0.67400,0.18800}%

\begin{tikzpicture}

\begin{axis}[%
width=4.521in,
height=3.522in,
at={(0.758in,0.525in)},
scale only axis,
xmin=-24,
xmax=0,
% *** CORREZIONE CRUCIALE QUI: "font=" deve precedere "\Large" ***
xlabel style={font=\Large\color{white!15!black},yshift=-5pt},
xlabel={SNR [dB]},
ymode=log,
ymin=20,
ymax=2367,
yminorticks=true,
% *** CORREZIONE CRUCIALE QUI: "font=" deve precedere "\Large" ***
ylabel style={font=\Large\color{white!15!black}},
ylabel={Weighted RMSE [Hz]},
axis background/.style={fill=white},
tick label style={font=\Large}, % Questa riga era già corretta
xmajorgrids,
ymajorgrids,
yminorgrids,
legend style={at={(0.02,0.02)}, anchor=south west, legend cell align=left, align=left, draw=white!15!black, font=\large}
]
\addplot [color=red, line width=2pt, mark=o, mark options={solid, red ,mark size=4pt}]
  table[row sep=crcr]{%
-24	1323.00471316849\\
-22	1243.92260154543\\
-20	1130.36817878345\\
-18	980.760829414945\\
-16	829.736455724945\\
-14	683.746047940347\\
-12	436.498972583228\\
-10	201.761728867548\\
-8	97.9465745350527\\
-6	84.4432378675807\\
-4	47.1655018397816\\
-2	32.479736198332\\
0	28.0062604988379\\
};
\addlegendentry{Zero Doppler}

\addplot [color=mycolor1, line width=2pt, mark=diamond, mark options={solid, mycolor1,mark size=4pt}]
  table[row sep=crcr]{%
-24	1367.5234947506\\
-22	1254.04416886086\\
-20	1070.58342253555\\
-18	868.929879848577\\
-16	710.015220576129\\
-14	557.83465030126\\
-12	322.657798492275\\
-10	151.675599127378\\
-8	82.2183056815932\\
-6	62.2846839141392\\
-4	47.1655931142258\\
-2	32.479736198332\\
0	28.0062604988379\\
};
\addlegendentry{EVM-based}

\addplot [color=mycolor2, line width=2pt, mark=square, mark options={solid, mycolor2,mark size=4pt}]
  table[row sep=crcr]{%
-24	1431.94624597946\\
-22	1400.99319983364\\
-20	1276.55143088822\\
-18	1104.03847956575\\
-16	884.405814527795\\
-14	660.586700204439\\
-12	384.219864532288\\
-10	142.908948536013\\
-8	81.6854484318888\\
-6	62.2846889123703\\
-4	47.1655931142258\\
-2	32.479736198332\\
0	28.0062604988379\\
};
\addlegendentry{DL-based}

\addplot [color=red, dashed, line width=2pt, mark=o, mark options={solid, red ,mark size=4pt}]
  table[row sep=crcr]{%
-24	1164.69812777668\\
-22	1157.47070604136\\
-20	1153.51536317893\\
-18	1155.40520442058\\
-16	1156.80782607119\\
-14	1156.23074601244\\
-12	1152.36959965585\\
-10	1163.81639225619\\
-8	1158.12298241946\\
-6	1151.96709566106\\
-4	1157.06315005796\\
-2	1165.65070300482\\
0	1151.38513173615\\
};
\addlegendentry{zD-based Initialization}

\addplot [color=mycolor1, dashed, line width=2pt, mark=diamond, mark options={solid, mycolor1,mark size=4pt}]
  table[row sep=crcr]{%
-24	1373.09416358858\\
-22	1271.36966158332\\
-20	1158.80957335234\\
-18	1079.17200029077\\
-16	1028.55997267745\\
-14	979.342464698439\\
-12	938.005600846624\\
-10	918.352598128624\\
-8	899.939945719631\\
-6	890.566831213284\\
-4	893.81691488886\\
-2	895.590828722463\\
0	886.014804568949\\
};
\addlegendentry{EVM-based Initialization}

\addplot [color=mycolor2, dashed, line width=2pt, mark=square, mark options={solid, mycolor2, mark size=4pt}]
  table[row sep=crcr]{%
-24	1628.46009284064\\
-22	1586.79956999597\\
-20	1554.93335603571\\
-18	1478.95057473423\\
-16	1404.32509467093\\
-14	1307.12503208705\\
-12	1174.31798060531\\
-10	1037.17007503064\\
-8	897.472598282643\\
-6	759.096509666555\\
-4	631.944903608546\\
-2	527.661233217673\\
0	436.466406864608\\
};
\addlegendentry{DL-based Initialization}

\begin{comment}
\addplot [color=black, dashed, line width=2pt]
  table[row sep=crcr]{%
-24	110.983719712792\\
-22	88.1575021625969\\
-20	70.0259930705176\\
-18	55.623623460519\\
-16	44.1834146323622\\
-14	35.0961337490153\\
-12	27.8778499665026\\
-10	22.1441633518004\\
-8	17.5897341846817\\
-6	13.9720225041876\\
-4	11.0983719712793\\
-2	8.81575021625969\\
0	7.00259930705176\\
};
\addlegendentry{Weighted CRLB}
\end{comment}

\addplot [color=black, dashed, line width=2pt]
 table[row sep=crcr]{%
-24	222.805252316496\\
-22	177.368590919755\\
-20	141.376076889674\\
-18	112.90963750795\\
-16	90.4507203341534\\
-14	72.7982878313862\\
-12	59.0028335733598\\
-10	48.3124432922942\\
-8	40.1276873170796\\
-6	33.963302932671\\
-4	29.4165214856852\\
-2	26.1439496283214\\
0	23.8491484729908\\
};
\addlegendentry{Weighted MCRLB}

\end{axis}

% Rimuovi o commenta questo blocco se non strettamente necessario
% per evitare problemi di capacità di TeX
% \begin{axis}[%
% width=5.833in,
% height=4.375in,
% at={(0in,0in)},
% scale only axis,
% xmin=0,
% xmax=1,
% ymin=0,
% ymax=1,
% axis line style={draw=none},
% ticks=none,
% axis x line*=bottom,
% axis y line*=left
% ]
% \end{axis}

\end{tikzpicture}%
}
    \caption{The RMSE is shown against SNR for the different initialization approaches and maximum UE speed of $300$ km/h. The MCRLB is shown as a theoretical limit for estimation performance. }\label{fig:rmseDoppler300}
    \end{subfigure}
    \hfill
    \begin{subfigure}[t]{0.31\textwidth}
        \centering
        \resizebox{0.98\columnwidth}{!}{
     \definecolor{mycolor1}{rgb}{0.49400,0.18400,0.55600}%
\definecolor{mycolor2}{rgb}{0.46600,0.67400,0.18800}%

\begin{tikzpicture}

\begin{axis}[%
width=4.521in,
height=3.522in,
at={(0.758in,0.525in)},
scale only axis,
xmin=-24,
xmax=0,
% *** CORREZIONE CRUCIALE QUI: "font=" deve precedere "\Large" ***
xlabel style={font=\Large\color{white!15!black},yshift=-5pt},
xlabel={SNR [dB]},
ymode=log,
ymin=20,
ymax=6367,
yminorticks=true,
% *** CORREZIONE CRUCIALE QUI: "font=" deve precedere "\Large" ***
ylabel style={font=\Large\color{white!15!black}},
ylabel={Weighted RMSE [Hz]},
axis background/.style={fill=white},
tick label style={font=\Large}, % Questa riga era già corretta
xmajorgrids,
ymajorgrids,
yminorgrids,
legend style={at={(0.02,0.02)}, anchor=south west, legend cell align=left, align=left, draw=white!15!black, font=\large}
]
\addplot [color=red, line width=2pt, mark=o, mark options={solid, red ,mark size=4pt}]
  table[row sep=crcr]{%
-24	4408.61363013322\\
-22	4296.4185256685\\
-20	4207.98131425138\\
-18	4111.35819847452\\
-16	3945.8082702561\\
-14	3757.16831761675\\
-12	3698.73933124199\\
-10	3784.97726258084\\
-8	3886.46524489324\\
-6	4013.09596245978\\
-4	4072.672169332\\
-2	4080.30395406221\\
0	3993.32916081715\\
};
\addlegendentry{Zero Doppler}

\addplot [color=mycolor1, line width=2pt, mark=diamond, mark options={solid, mycolor1,mark size=4pt}]
  table[row sep=crcr]{%
-24	4080.61851132243\\
-22	3857.87019373624\\
-20	3642.45461222528\\
-18	3385.25284804886\\
-16	3106.10814654021\\
-14	2770.76509787521\\
-12	2571.27408175021\\
-10	2449.89349389897\\
-8	2489.27932239864\\
-6	2592.50123268303\\
-4	2622.7152328362\\
-2	2687.56531510364\\
0	2662.24215627798\\
};
\addlegendentry{EVM-based}

\addplot [color=mycolor2, line width=2pt, mark=square, mark options={solid, mycolor2,mark size=4pt}]
  table[row sep=crcr]{%
-24	4293.32484268529\\
-22	4094.43909368079\\
-20	3812.55169306258\\
-18	3401.64421671539\\
-16	2705.70575468487\\
-14	1945.86210699016\\
-12	1171.45835090998\\
-10	548.100988690729\\
-8	311.829829481187\\
-6	191.345493141583\\
-4	147.563140784395\\
-2	120.714993826014\\
0	100.209196515246\\
};
\addlegendentry{DL-based}

\addplot [color=red, dashed, line width=2pt, mark=o, mark options={solid, red ,mark size=4pt}]
  table[row sep=crcr]{%
-24	3864.42359626391\\
-22	3861.1489589711\\
-20	3859.82290646329\\
-18	3865.45230249557\\
-16	3870.85314579693\\
-14	3855.2421292602\\
-12	3868.03573038376\\
-10	3866.96188543535\\
-8	3859.71186874362\\
-6	3855.72907869767\\
-4	3889.19054087818\\
-2	3875.05067937564\\
0	3854.43043530535\\
};
\addlegendentry{zD-based Initialization}

\addplot [color=mycolor1, dashed, line width=2pt, mark=diamond, mark options={solid, mycolor1,mark size=4pt}]
  table[row sep=crcr]{%
-24	3524.24093036007\\
-22	3320.3344565817\\
-20	3187.85312798044\\
-18	3081.17052258918\\
-16	3030.0655633643\\
-14	2981.78386329656\\
-12	2979.98978534634\\
-10	2978.15956632696\\
-8	2970.0832079762\\
-6	2966.03892322459\\
-4	2997.97359078371\\
-2	2985.59930157178\\
0	2968.11500693115\\
};
\addlegendentry{EVM-based Initialization}

\addplot [color=mycolor2, dashed, line width=2pt, mark=square, mark options={solid, mycolor2, mark size=4pt}]
  table[row sep=crcr]{%
-24	3781.99367709722\\
-22	3662.842553791\\
-20	3453.06528631055\\
-18	3175.62827526067\\
-16	2841.92894737396\\
-14	2453.55518159249\\
-12	2093.8797739215\\
-10	1738.31508565523\\
-8	1453.24536842606\\
-6	1192.47377435901\\
-4	982.808503165115\\
-2	813.043753954657\\
0	678.617204105763\\
};
\addlegendentry{DL-based Initialization}

\addplot [color=black, dashed, line width=2pt]
 table[row sep=crcr]{%
-24	222.805252316496\\
-22	177.368590919755\\
-20	141.376076889674\\
-18	112.90963750795\\
-16	90.4507203341534\\
-14	72.7982878313862\\
-12	59.0028335733598\\
-10	48.3124432922942\\
-8	40.1276873170796\\
-6	33.963302932671\\
-4	29.4165214856852\\
-2	26.1439496283214\\
0	23.8491484729908\\
};
\addlegendentry{Weighted MCRLB}

\end{axis}

% Rimuovi o commenta questo blocco se non strettamente necessario
% per evitare problemi di capacità di TeX
% \begin{axis}[%
% width=5.833in,
% height=4.375in,
% at={(0in,0in)},
% scale only axis,
% xmin=0,
% xmax=1,
% ymin=0,
% ymax=1,
% axis line style={draw=none},
% ticks=none,
% axis x line*=bottom,
% axis y line*=left
% ]
% \end{axis}

\end{tikzpicture}%
}
    \caption{The RMSE is shown against SNR for the different initialization approaches and maximum UE speed of $1000$ km/h. The MCRLB is shown as a theoretical limit for estimation performance. }\label{fig:rmseDoppler1000}
    \end{subfigure}
    
    \caption{Simulation results for varying Doppler, SNR, and UE speed.} \vspace{-5mm}
\end{figure*}

\subsection{Resource Allocation and Utilization}

\subsubsection{Pilot Overhead}
A single \gls{OFDM} block is used for estimating channel parameters, thus the effective pilot overhead is $1/N$. Assuming a geometric coherence time of $10$ ms\footnote{In 6G high-mobility scenarios, the geometric coherence time can be in the order of tens of milliseconds \cite{Viterbo2022}, depending on the velocity of mobile scatterers.}, an \gls{OFDM} symbol duration of $T=33.3 \ \mu$s with a subcarrier spacing of $\Delta f=30$ kHz and a \gls{CP} duration of $5 \ \mu$s, the minimum overhead for a continuous transmission is about $0.38\%$. However, for a short frame duration of $1$ ms, the overhead is less than $4\%$.

\subsubsection{Complexity}
The complexity of the proposed receiver is dominated by the joint Doppler estimation and data detection procedure with a sliding buffer, thus the complexity is $\mathcal{O}\big(NKPMN_r)+\mathcal{O}(NKPM\log(M))\big)$. The \gls{EVM}- and \gls{DL}-based Doppler initialization modules have complexities of $\mathcal{O}(M\log(M))$ and $\mathcal{O}(M^2)$, respectively.

\subsubsection{Decoding Latency}
The sliding buffer used for Doppler refinement in Algorithm \ref{DOAaidedRXAlg} introduces a latency in decoding \gls{OFDM} symbols of $KT'$.

\begin{table}[t]
\centering
\caption{Simulation parameters.}\label{TabSim}
{
\begin{tabular}{l | l}
\hline
\multicolumn{2}{c}{\textbf{General}} \\
\hline
Carrier frequency, $f_c$ & $5.9$ GHz \\
\hline
Number of antennas, $N_r$ & $32$ \\
\hline
Number of subcarriers, $M$ & $128$ \\
\hline
Number of \gls{OFDM} symbols, $N$ & $32$ \\
\hline
Subcarrier spacing, $\Delta f$ & $30$ kHz \\
\hline
\gls{CP} duration, $T_{\text{CP}}$ & $5 \ \mu$s \\
\hline
Modulation & $4$-\gls{QAM} \\
\hline
Sliding window dimension, $K$ & $\min(\lfloor{1+\frac{1}{2\sigma_{\nu}T'}}\rfloor,N/2)$ \\
\hline
\multicolumn{2}{c}{\textbf{Wireless channel}} \\
\hline
Number of multipaths, $P$ & $4$ \\
\hline
Direction of arrivals, $\theta_p$ & $[10^\circ,50^\circ, -30^\circ,20^\circ]$ \\
\hline
Propagation delays, $\tau_p$ & $[0, 0.9, 2.4, 3] \ \mu$s \\
\hline
Average power per path, $\mathcal{P}_p$ & $[0, -1 , -5, -7]$ dB \\
\hline
Doppler shifts, $\nu_p$ & \makecell[l]{$\nu_p=f_c\frac{v_{\max}}{c}\cos(\theta)$ \\ $\theta\sim\mathcal{U}[0,2\pi]$} \\
\hline
\end{tabular}
}
\end{table}
 
\section{Simulation Results}
Numerical simulations are performed to validate the proposed approach. Table \ref{TabSim} summarizes the simulation parameters.
The \gls{FNN} is composed of four layers with dimensions $\big\{M,M,\frac{M}{2},\frac{M}{2}\big\}$. The layers are fully connected, and each layer employs a \gls{ReLU} as the activation function, except for the last layer, which adopts a linear activation in order to provide real numbers as output. The \gls{FNN} is trained with $5\cdot10^5$ pilot samples ($80\%$ training, $20\%$ validation) affected by random delay, Doppler, and \gls{SNR}. In particular, the delay and Doppler are uniformly distributed as $\tau\sim\mathcal{U}[0,5] \ \mu$s and $\nu\sim\mathcal{U}[-5,5]$ kHz, and \gls{SNR}$\sim\mathcal{U}[12,18]$ dB. The pilot samples are synthetically generated using $\mathbf{F}_N^{\mathsf{H}}\big(\mathbf{x}_1\odot\mathbf{b}(\tau)\big)\odot\mathbf{c}(\nu)+\mathbf{w}$, where $\mathbf{w}\sim\mathcal{CN}(\mathbf{0},\mathbf{I}/\text{SNR})$. Figure \ref{fig:testDoppler} shows the predicted Dopplers against true values, and it can be noted that the \gls{DL} architecture is actually capable of estimating the Doppler from the received block-type pilot.

Figure \ref{fig:bervssnr300} and Figure \ref{fig:bervssnr1000} show the \gls{BER} performance against \gls{SNR} for maximum speeds of $300$ km/h and $1000$ km/h, respectively. It can be noted that, in both scenarios, the proposed \gls{DoA}-aided receiver with perfect \gls{CSI} achieves the same performance as \gls{OFDM} over an \gls{AWGN} channel with coherent integration. For a $4$-\gls{QAM}, this lower bound can be approximated as $Q(\sqrt{N_r\text{SNR}})$, where $Q(\cdot)$ denotes the Gaussian Q-function. Moreover, when considering imperfect \gls{CSI}, the \gls{DL}-based Doppler initialization achieves \gls{BER} performance close to the \gls{AWGN} lower bound at sufficient \gls{SNR} values. Conversely, the \gls{zD}- and \gls{EVM}-based initialization schemes underperform the \gls{DL}-based approach and, in the case of a maximum speed of $1000$ km/h, they fail completely. Figure \ref{fig:bervsspeed} shows the \gls{BER} performance against the maximum speed. It can be noted that, unlike the cases of \gls{zD}- and \gls{EVM}-based initialization, which start failing at speeds higher than $300$ km/h and $400$ km/h respectively, the \gls{DL}-based approach achieves an almost constant \gls{BER} up to the maximum 6G speed of $1000$ km/h. Accordingly, Figure \ref{fig:rmseDoppler300} and Figure \ref{fig:rmseDoppler1000} show the Doppler weighted \gls{RMSE}, defined as
\begin{equation}
\text{Weighted RMSE}=\sqrt{\mathbb{E}\Bigg[\frac{\sum_{p=1}^P|\alpha_p|^2(\nu_p-\hat\nu_p)^2}{\|\boldsymbol{\alpha}\|^2}\Bigg]},
\end{equation}
against \gls{SNR} for maximum speeds of $300$ km/h and $1000$ km/h, respectively. It can be noted that the \gls{DL}-based initialization effectively provides more accurate coarse estimates than \gls{EVM}-based initialization. Moreover, the proposed Doppler estimation via gain tracking achieves performance sufficiently close to the \gls{MCRLB} for all the initialization approaches when the maximum speed is $300$ km/h. Conversely, in the case of a maximum speed of $1000$ km/h, only the \gls{DL}-based approach reaches low \gls{RMSE}.

\section{Conclusions and Future Works}
In this work, a multi-antenna \gls{RX}, empowered by \gls{DL}, for \gls{SIMO}-\gls{OFDM} is developed. A mobility-resilience numerical analysis is carried out to validate the proposed scheme. Simulation results reveal that the proposed \gls{DL}-based approach achieves almost constant \gls{BER} up to a maximum speed of $1000$ km/h with low pilot overhead (less than $4\%$) and low complexity inherited from simple operations (single-tap delay compensation and \gls{ICI} cancellation, \gls{DFT}, and gain tracking) at the cost of introducing a decoding latency due to the sliding buffer, making \gls{DoA}-aided receivers suitable for 6G high-mobility communications.
Future research will further investigate the impact of higher-order \gls{QAM} constellations and the effect of multiple paths with the same angle, which results in higher \gls{IPI}.

%\appendix{Appendix A}
\appendix[Modified Cramér-Rao Lower Bound for Doppler]
The weighted \gls{MCRLB} for the Doppler shift is computed as follows. Assuming small \gls{IPI} due to large \gls{ULA} at the \gls{RX}, the \gls{CFI} at the $n$-th \gls{OFDM} symbol is approximated using the Slepian-Bangs formula \cite{Kay1993} as 
\begin{equation} \label{eq:SlepianBangs}
\mathcal{I}_{p,n}(\nu_p|\alpha_p, \mathbf{x}_n)\approx \frac{2}{\sigma^2+\mathcal{P}_{\text{IPI}}^{(p)}}\Bigg\|\frac{\partial\boldsymbol{\mu}_p}{\partial\nu_p}\Bigg\|^2,
\end{equation}
where $\boldsymbol{\mu}_p=\sqrt{P_T}\alpha_p e^{j2\pi \nu_pt_n}\big[\mathbf{F}_{M}^{\mathsf{H}}\big(\mathbf{x}_n\odot\mathbf{b}(\tau_p)\big)\odot\mathbf{c}(\nu_p)\big]$ from \eqref{eq:AngleDomainMF} and $\mathcal{P}_{\text{IPI}}^{(p)}=P_T|\alpha_p|^2\sum_{i=1}^P(1-\delta_{ip})|\mathbf{a}^\top(\theta_p)\mathbf{a}^*(\theta_i)|^2$. Therefore, the derivative is 
\begin{equation}\label{eq:SlepianBangs}
\Big[\frac{\partial\boldsymbol{\mu}_p}{\partial\nu_p}\Big]_q= 
\frac{j2\pi(t_n+q\Delta\tau)\sqrt{P_T}\alpha_p}{e^{-j2\pi\nu_p(t_n+q\Delta\tau)}}\Big[\mathbf{F}_{M}^{\mathsf{H}}\big(\mathbf{x}_n\odot\mathbf{b}(\tau_p)\big)\Big]_q.
\end{equation}
Thus, the \gls{CFI} becomes
\begin{equation}\begin{split}\label{eq:FIM_n}
\mathcal{I}_{p,n}(\nu_p|\alpha_p,\mathbf{x}_n)\approx & \\ \frac{8\pi^2|\alpha_p|^2P_T}{\sigma^2+\mathcal{P}_{\text{IPI}}^{(p)}} & \sum_{q=0}^{M-1}(t_n+q\Delta\tau)^2\Big|\big[\mathbf{F}_{M}^{\mathsf{H}}\big(\mathbf{x}_n\odot\mathbf{b}(\tau_p)\big)\big]_q\Big|^2.
\end{split}\end{equation}
In particular, for $n=1$ (deterministic pilot)
\begin{equation} \label{eq:FIM_pilot}
\mathcal{I}_{p,1}(\nu_p|\alpha_p)\approx\frac{8\pi^2|\alpha_p|^2P_T}{\sigma^2+\mathcal{P}_{IPI}^{(p)}}\sum_{q=0}^{M-1}(T_{\text{CP}}+q\Delta\tau)^2\Big|\big[\mathbf{\tilde{b}}(\tau_p)\big]_q\Big|^2.
\end{equation}
The aggregated averaged information comprising the $N$ received observations is given as
\begin{equation}\begin{split} \label{eq:fullFIM}
\mathcal{I}_p(\nu_p)&=\mathbb{E}_{\alpha}\big[\mathcal{I}_{p,1}(\nu_p|\alpha_p)\big]+\sum_{n=2}^N\mathbb{E}_{\alpha,\mathbf{x}}\big[\mathcal{I}_{p,n}(\nu_p|\alpha_p,\mathbf{x}_n)\big]\approx
\\
\frac{8\pi^2\mathcal{P}_pP_T}{\sigma^2+\mathcal{P}_{\text{IPI}}^{(p)}}&\Big(\sum_{q=0}^{M-1}(T_{\text{CP}}+q\Delta\tau)^2\Big|\big[\mathbf{\tilde{b}}(\tau_p)\big]_q\Big|^2+\|\boldsymbol{\Theta}\|^2\Big),
\end{split}\end{equation}
where $\mathcal{P}_p=\mathbb{E}[|\alpha_p|^2]$ is the average power of the $p$-th path and $\boldsymbol{\Theta}\in\mathbb{R}^{(N-1)M}$ where $[\boldsymbol{\Theta}]_i=t_n+q\Delta\tau$ with $i=M(n-2)+q$ where $n=2,\dots,N$ and $q=0,\dots,M-1$. The \gls{MCRLB} for the $p$-th path is then obtained as $\text{MCRLB}_p=\mathcal{I}^{-1}_p(\nu_p)$. Hence, the global weighted bound is  %$\text{Weighted MCRLB}={\sum_{p=1}^P\mathcal{P}_p\text{MCRLB}_p}/{(\sum_{p=1}^P\mathcal{P}_p)}$.
\begin{equation}
\text{Weighted MCRLB}=\frac{\sum_{p=1}^P\mathcal{P}_p\text{MCRLB}_p}{\sum_{p=1}^P\mathcal{P}_p}.
\end{equation}

\ifCLASSOPTIONcaptionsoff
  \newpage
\fi

\balance
\bibliographystyle{IEEEtran}
\bibliography{reference}

\end{document}